%Paper: hep-th/9506204
%From: dowker@a3.ph.man.ac.uk
%Date: Fri, 30 Jun 1995 14:41:51 BST
%Date (revised): Mon, 03 Jul 1995 09:30:53 BST

\magnification=1200
\font\open=msbm10 %scaled\magstep1 % For VAX. Borde p195.
\font\opens=msbm8 %scaled\magstep1 % For VAX. Borde p195.
%\font\open=msym10 %scaled\magstep1 % For Arbortxt on PC
%\font\opens=msym8 %scaled\magstep1 % For Arbortxt on PC
%\font\goth=eufm10  % For Arbortxt on PC and VAX. Borde p199
\def\mbox#1{{\leavevmode\hbox{#1}}}
\def\hspace#1{{\phantom{\mbox#1}}}
\def\oR{\mbox{\open\char82}}
\def\osR{\mbox{\opens\char82}}

\def\rS{{\rm S}}
\def\rHS{{\rm HS}}

\def\al{\alpha}

\def\de{\delta}
\def\Ga{\Gamma}

\def\ep{\epsilon}

\def\ka{\kappa}
\def\la{\lambda}

\def\om{\omega}

\def\si{\sigma}

\def\th{\theta}
\def\Th{\Theta}
\def\ze{\zeta}
\def\De{\Delta}

\def\Det{{\rm Det\,}}
\def\Real{{\rm Re\,}}

\def\slice{{\rm slice}}
\def\cylinder{{\rm cylinder}}
\def\shell{{\rm shell}}
\def\ball{{\rm ball}}

\def\zf{$\zeta$--function}
\def\zfs{$\zeta$--functions}

     % Newline

\def\frac#1/#2{\leavevmode\kern.1em
\raise.5ex\hbox{\the\scriptfont0 #1}\kern-.1em/\kern-.15em
\lower.25ex\hbox{\the\scriptfont0 #2}}
\def\sfrac#1/#2{\leavevmode\kern.1em
\raise.5ex\hbox{\the\scriptscriptfont0 #1}\kern-.1em/\kern-.15em
\lower.25ex\hbox{\the\scriptscriptfont0 #2}}

\def\ds{{|\!|}}        %double stroke
\def\cd#1{{}_{\ds #1}} %lower covariant deriv.
    %lower ordinary  deriv.
\def\noin{\noindent}

\def\comb#1#2{{\left(#1\atop#2\right)}}

\def\cosech{{\rm cosech\,}}
\def\etc{{\it etc. }}

\def\pa{\partial}

\def\tr{{\rm tr\,}}

\def\man{{\cal M}}

\def\fd{{\cal F}}
\def\wR{{\widehat R}}
\rightline{Moscow 1995}
\vskip 5truept
\centerline{\bf FUNCTIONAL DETERMINANTS}
\centerline{\bf ON CERTAIN DOMAINS}
\vskip 15truept
\centerline{J.S.Dowker and J.S.Apps}
\vskip10truept
\centerline{\it Department of Theoretical Physics,}
\centerline{\it The University of Manchester,}
\centerline{\it Manchester, England.}
\vskip20truept
\noin {\bf Introduction}

In this talk I would just like to report on some technical
calculations of functional determinants that we have been amusing ourselves
with in Manchester.

The set up is an elliptic, positive second order operator $D$, which will
normally be $-\De_2+\xi R$ with $\xi$ chosen to give a conformally invariant
theory, defined on a Riemannian manifold $\man$ with a boundary that may be
non-smooth. Some massless spin-half results will also be listed. No
motivation will be presented.

The functional determinant is defined, as usual, by
$$
\ln\Det D=-\ze'(0), \quad W\equiv{1\over2}\ln\Det D,
$$ where $W$ is, in field theory terms, a one-loop effective action.

I would like to distinguish two methods of deriving $W$:
\vskip10truept
\noindent A. Conformal methods
\vskip10truept
\noindent and
\vskip10truept
\noindent B. Direct methods.
\vskip10truept
Conformal methods are only easily implemented for conformally invariant
situations, and then just give the difference in $W$'s for conformally
related spaces. To obtain $W$ in one space, a direct calculation, by which I
mean one from the definition using eigenvalue properties, still has to be
performed.

Since this is not meant to be a review talk, in place of systematic references
I simply list the names of some of the workers who have contributed to this
area of investigation:
\vskip10truept
\noindent Aurell, Barvinsky, Bordag, Branson, D'Eath, Elizalde, Esposito,
Frolov, Gilkey, Kamenshchik, Karmazin, Kennedy, Kirsten, Louko, Mishakov,
Moss, Pollifrone, Salomonson, Schleich, Vardi, Vassilevich, Voros, Weisberger,
Wipf, Zelnikov $\ldots$
\vskip10truept
\noindent However, a few historical remarks will be made at the end.
\vskip15truept
\noin {\bf A. Conformal method.}

By integrating the conformal anomaly (which equals $\ze(0)$ and is, more or
less, the constant term in the heat-kernel short-time expansion) between two
conformally related metrics $g\to\bar g=\exp(-2\om)g$, one obtains in a
standard fashion,
$${1\over2}\ln{\Det\overline D\over\Det D}=W[\bar g,g].
$$

In two dimensions the explicit form of $W[\bar g,g]$ is familiar following
the work of L\"uscher, Symanzik and Weiss, of Polyakov and of Alvarez. I will
not consider this case but concentrate on three and four dimensions.

In three dimensions, for Dirichlet (D) conditions
$$
W^D[\bar g,g]={1\over1536\pi}\int_{\pa\man}\bigg[\big(6\tr(\ka^2)-3\ka^2
-16{\widehat\De}_2\om-4\widehat R+30\ka N
+18N^2-24n^\mu n^\nu \om_{\mu\nu}\bigg],
$$ where
$N=n^\mu\pa_\mu\om$. This has only a boundary part.
(In this talk, I exhibit only the Dircihlet forms, but the Robin case
is available as well.)

\noindent In four dimensions
$$\eqalign{
W^D[\bar g,g]=&{1\over2880\pi^2}\int_\man\bigg[(|{\rm Riem}|^2-
|{\rm Ric}|^2+\De_2R)\om-2R_{\mu\nu}\om^\mu\om^\nu-4\om^\mu\om_\mu\De_2\om\cr
&\hspace{********************}+2(\om^\mu\om_\mu)^2+3(\De_2\om)^2\bigg]\cr
&+{1\over5760\pi^2}\int_{\pa\man}\bigg[
\bigg({320\over21}\tr(\ka^3)-{88\over7}\ka\tr(\ka^2)
+{40\over21}\ka^3-4R_{\mu\nu}\chi^{\mu\nu}-4\ka R_{\mu\nu}n^\mu n^\nu\cr
&+16R_{\mu\nu\rho\si}n^\mu n^\rho\chi^{\nu\si}-
2n^\mu\pa_\mu R\bigg)
-N\bigg({12\over7}\ka^2-{60\over7}\tr(\ka^2)-12\De_2\om+8\om^\mu\om_\mu
\bigg)\cr
&-\!{4\over7}N^2\ka\!+\!{16\over21}N^3\!+\!24\ka\De_2\om\!-\!4\chi^
{\mu\nu}\om_\mu\om_\nu-
20\ka\om^\mu\om_\mu\!-\!30n^\mu\pa_\mu(\De_2\om\!-\!\om^\nu\om_\nu)\bigg].\cr}
$$
The volume part of this was first evaluated by Brown and by Riegert in 1984
but the boundary part had to wait upon the complete heat-kernel coefficient.
\vskip15truept
\noin{\bf Conformal transformations.}

It is simply a game to find interesting manifolds on which to use these
expressions. There are clearly a large number available but
the ones I have selected are associated with the stereographic projection of
the sphere onto the (equatorial) plane, and thence to the cylinder. In fact
I shall also want to consider parts of spheres such as the hemisphere.

In symbols the mappings are
$$\rS^{d+1}\longleftrightarrow\oR^{d+1}(\sim\oR^+\times\rS^d)
\longleftrightarrow\oR\times\rS^d.$$
My strategy is based on the fact that it seems easier to calculate
the functional determinant on the (hemi-)sphere and the cylinder, than on
the Euclidean ball or spherical cap.

In terms of metrics, the equatorial stereographic
projection, $\rS^{d+1}\to\oR^{d+1}$ is
$$
d\si^2_{d+1}={4\over(1+r^2)^2}d{\bf r}^2.
$$
For example this maps the hemisphere onto the equatorial disc (or ball).
Then, after rescaling the ball, an inverse projection yields a spherical
{\it cap}.

The map between a Euclidean spherical shell and a cylinder is
$$\oR\times\rS^d\longrightarrow\oR^{d+1}$$
$$
ds^2=d\tau^2+d\si^2_{d}=e^{-2\tau}\big(dr^2+r^2d\si^2_{d}\big),
\quad\quad r=\exp\tau.$$
An inverse stereographic projection takes such a shell to a
{\it slice} of the sphere, $\rS^{d+1}$.
\vskip15truept
\noin{\bf Caps and balls.}

Having briefly indicated the conformal transformations, all that is left is the
routine computation of the necessary geometrical quantities -- extrinsic
curvatures \etc -- on the appropriate domains and the calculation of
$W[\bar g,g]$.

I just list the results, because that is all I have advertised. If you wish
to see more details, and a few graphs, then the paper in
{\it Class.Quant.Grav.}{\bf12}(1995)1363 can be consulted.

$$
W^D_{3\ball}=W^D_{3\rm hemisphere}+{7\over64},
$$
$$
W^D_{\rm 3cap}(\th)-W^D_{\rm 3hemisphere}={1\over48}\big(\ln\sin\th+
{21\over4}
\cos^2\th\big),
$$
$$
W^D_{4\ball}=W^D_{4\rm hemisphere}-{1\over180}\ln2-{17\over15120},
$$
$$
W^D_{\rm4cap}(\th)-W^D_{\rm 4hemisphere}=
{1\over180}\bigg({1\over168}\big(1365\cos\th-1399\cos^3\th\big)
\ln\tan\th/2\bigg),
$$
where
$$
W^D_{\rm
3hemisphere}={3\over8}\ze'_R(-2)-{1\over4}\ze_R'(-1)-{1\over16}+
{1\over24}\ln2.
$$and
$$
W^D_{\rm 4hemisphere}=-{1\over6}\ze'_R(-3)+{1\over4}\ze'_R(-2)
-{1\over12}\ze_R'(-1)-{1\over516}.
$$

The direct computation of these hemisphere values is another story. Some
related calculations will be outlined later.
\vskip15truept
\noin{\bf Shells and slices.}

Weisberger has shown that, in two dimensions,
$$
W^D_{2\shell}=W^D_{I\times\rS^1}-{1\over12}\ln\big(r_1/r_2\big)
$$
($r_1$ and $r_2$ = the outer and inner radii of the shell.)

Our higher-dimensional results are
$$
W^D_{3\shell}=W^D_{I\times\rS^2}+{1\over96}\big(2\ln(r_1 r_2)+9\big),
$$
$$
W^D_{4\shell}=W^D_{I\times\rS^3}+{1\over720}\ln\big(r_1/r_2\big),
$$
$$
W^D_{2\slice}=W^D_{2\shell}+{2\over3}\sin\Th\sin\De,
$$
$$
W^D_{3\slice}=W^D_{3\shell}-{7\over64}\big(1-\cos2\Th\cos2\De\big)
+{1\over24}\ln\big(\cos\De+\cos\Th\big),
$$
$$
W^D_{4\slice}=W^D_{4\shell}-{1\over30240}\sin\Th\sin\De\,\big(2730
-1399(2+2\cos2\Th\cos2\De+\cos2\Th+\cos2\De)\big),
$$where
$\Th$ is the colatitude of the midpoint of the slice, and $\De$ is its
angular half-width.
\vskip15truept
\noin {\bf Spin-half.}

I now give the (massless) spin-half results as some of these have not been
exhibited before. They were calculated by Jonathan Apps and contain exactly
the same ingredients. A new $W[\bar g,g]$ was found from the smeared versions
of the unsmeared heat-kernel coefficients given by Moss and Poletti, for
mixed boundary conditions. A new evaluation of the spinor determinants on a
hemisphere was also performed. The conclusions are as follows,
$$
W_{2\ball}=W_{2\rm hemisphere}-{1\over12}+{1\over6}\ln2,
$$
$$
W_{3\ball}=W_{3\rm hemisphere}+{1\over16},
$$
$$
W_{4\ball}=W_{4\rm hemisphere}+{1\over720}\bigg({251\over21}
-44\ln2\bigg),
$$
$$
W_{\rm2cap}(\th)-W_{\rm 2hemisphere}=-{1\over12}\cos\th-{1\over6}\ln\tan\th/2,
$$
$$
W_{\rm3cap}(\th)-W_{\rm 3hemisphere}={1\over16}\cos^2\th,
$$
$$
W_{\rm4cap}(\th)-W_{\rm 4hemisphere}
={1\over180}\bigg({1\over168}\big(1155\cos\th
-653\cos^3\th\big)+11\ln\tan\th/2\bigg),
$$
where
$$
W_{\rm 2hemisphere}=2\ze'_R(-1),
$$
$$
W_{\rm 3hemisphere}=-{3\over2}\ze'_R(-2)+{1\over4}\ln2,
$$
$$
W_{\rm 4hemisphere}={2\over3}\ze'_R(-3)-{2\over3}\ze_R'(-1).
$$

Perhaps I should point out that in an $N$-dimensional space, the spinor
dimensions are $2^{N/2}$ if $N$ is odd and $2^{(N+1)/2}$ if $N$ is even.
\vfill\eject
\noin{\bf Generalised cylinders.}

In order to complete the evaluation of the determinants on slices and shells,
those on cylinders have to be determined. Instead of recounting results I will
now give some formalism. This is a direct calculation in the case that
the eigenvalues are known explicitly.
The Dirichlet \zf\ on $I\times\man$, is
$$\ze^D_{I\times\man}(s)=\sum_{n=1}^\infty\sum_\la{d_\la\over
(\pi^2n^2/L^2+\la^2)^s}.
$$  $L$ is the length of the interval $I$. $d_\la$
is the degeneracy of
the $\la$ eigenvalue of the scalar Laplacian $-\De_2+\xi R$,
$\xi=(d-1)/4d$, on $\man$, conformal in $(d+1)$-dimensions.

It is convenient to turn the interval into a circle and write,
$$
\ze_{I\times\man}(s)={1\over2}\big(\ze_{S^1\times\man}(s)\mp\ze_\man(s)
\big)
$$
which has a thermal look about it.
The signs refer to the boundary conditions {\it on the interval}.

A standard transformation, analogous to that that gives the Kronecker limit
formula, gives
$$
\ze_{I\times\man}'(0)
={1\over2}\big(2L\ze_{\osR\times\man}'(0)\mp
\ze_\man'(0)\big)+\sum_{m=1}^\infty{1\over m}K^{1/2}(2mL).
$$
The first term corresponds to the zero temperature free energy (vacuum energy)
and the summation to the standard statistical mechanical mode sum.

$K^{1/2}(z)$ is the kernel for $(-\De_2+\xi R)^{1/2}$.
and
$$
\ze_{\osR\times\man}(s)\equiv{1\over\sqrt{4\pi}}{\Ga(s-1/2)
\over\Ga(s)}\ze_\man(s-1/2).
$$

The cylinder in which we are interested has the sphere $\rS^d$ as
cross-section, $\man$. However allow me to be a little more general and
take $\man$ to be $\fd$, the fundamental domain on $\rS^d$ for the complete
symmetry group of a regular $(d+1)$--polytope . This is a finite reflection
group generated by reflections in $d+1$ hyperplanes passing through the origin
in the ambient $\oR^{d+1}$ and is classified by the degrees $d_i$,
$(i=1,2,\ldots,d_{d+1}=2)$ of the algebraically independent invariant
polynomials in the cartesian coordinates of the $\oR^{d+1}$. When $d=2$,
$\fd$ is a geodesic right-spherical triangle and for $d=3$ it is a spherical,
rectangular, geodesic tetrahedron familiar since the time of Lobachevski and
Schl\"afli.

The Neumann and Dirichlet  \zfs\ on $\fd$ can be determined essentially by
images, group theory and invariant theory. They are given by a Barnes \zf,
$$
\ze_\fd(s)=\ze_d\left(2s,a\mid {\bf d}\right),
$$
whose definition is
$$\eqalign{
\zeta_d(s,a|{\bf {d}})&=\!{i\Ga(1-s)\over2\pi}\!\!\int_L\! dz {\exp(-a z)
(-z)^{s-1}\over\prod_{i=1}^d\!\big(1\!-\exp(-d_i z)\big)}\cr
&=\sum_{{\bf {m}}={\bf 0}}^\infty{1\over(a+{\bf {m.d}})^s},\qquad
\Real\, s>d.\cr}
$$
This is a generalisation of the Hurwitz \zf\ which has just one factor in the
denominator. The eigenvalues are
$$
\la_n=(a+{\bf m.d})^2,
$$
and the constant $a$ is $(d-1)/2$ for Neumann and  $\sum d_i-(d-1)/2$
for Dirichlet conditions.

Using
$$
\ze_{\osR\times\fd}'(0)=-\ze_\fd(-1/2),
$$
the Kronecker limit formula becomes
$$
\ze_{I\times\fd}'(0)=-L\ze_d\big(-1,a\mid {\bf d}\big)
\mp\ze_d'\big(0,a\mid {\bf d}\big)+\sum_{m=1}^\infty{e^{-2amL}
\over m}\prod_{i=1}^d{1\over1-q_i^m}.
$$ where $q_i=\exp(-2Ld_i)$. Barnes gives, for example,
$$\ze_d\big(-1,a\mid {\bf
d}\big)={(-1)^d\over\prod_id_i}{1\over(d-1)!}B^{(d)}_{d+1}
\big(a\mid{\bf d}\big).
$$\
and,in this way, one can proceed to give a general treatment. However I shall
consider the special case of the hemisphere which, for $d=2$, is a
geodesic triangle with all angles equal to $\pi$. Every degree, $d_i$, is
unity and the Barnes \zf\ reduces to
$$\eqalign{
\ze_d(s,a)&={i\Ga(1-s)\over2\pi}\int_L{e^{z(d/2-a)}(-z)^{s-1}\over
2^d\sinh^d(z/2)}\,dz\cr
&=\sum_{m=0}^\infty\comb{m+d-1}{d-1}{1\over{(a+m)}^s}\,,\cr
}
$$ with $a=(d+1)/2$ for Dirichlet and $a=(d-1)/2$ for
Neumann conditions {\it on the hemisphere rim}. For completeness I indicate the
further progress of this direct method.

The summations can be manipulated into
$$
\ze^D_d(s)={1\over(d-1)!}\sum_{m=1}^\infty{(m+q-1)\ldots(m-q)\over m^s}
$$
for odd $d=2q+1$ and into
$$
\ze^D_d(s)={1\over(d-1)!}\sum_{m=0}^\infty{(m+q)\ldots(m-q)
\over(m+1/2)^s},
$$
for even $d=2q+2$.

As usual, the numerators are expanded in Stirling numbers
$$\eqalign{
&(m+a)(m+a-1)\ldots(m+a-b+1)\cr
&=\sum_{k=0}^bS^{(k)}(a,b)\,m^k=\sum_{k=0}^bT^{(k)}(a,b)\,(m+1/2)^k.\cr}
$$
in order to give a finite series of standard \zfs,
$$
\ze^D_d(s)=\sum_{k=0}^{2q}S^{(k)}(q-1,2q)\,\ze_R(s-k),
$$
for odd $d$ and
$$
\ze^D_d(s)=\sum_{k=0}^{2q+1}T^{(k)}(q,2q+1)\,\ze_R(s-k,1/2),
$$
for even $d$.

Some specific cases are
$$
\ze'_{I\times\rHS^2}(0)=
\bigg({1\over2}\ze_R'(-1)+{1\over24}(1\mp6)\ln2\bigg)\mp{L\over48}
-{1\over4}\sum_{m=1}^\infty{e^{\pm mL}\over m\sinh^2\!mL},
$$
$$
\ze'_{I\times\rHS^3}(0)=-\bigg({1\over2}\ze_R'(-2)\mp{1\over2}
\ze_R'(-1)\bigg)-{L\over240}+{1\over8}\sum_{m=1}^\infty{e^{\pm mL}
\over m\sinh^3\!mL},
$$

$$
W^{D,N}_{I\times\rS^2}=\mp{1\over2}\ze'_R(-1)\mp{1\over24}\ln2
-{1\over4}\sum_{m=1}^\infty{\cosh mL\over m\sinh^2\!mL},
$$
$$
W^{D,N}_{I\times\rS^3}=\pm{1\over2}\ze'_R(-2)+{L\over240}-{1\over8}
\sum_{m=1}^\infty{\cosh mL\over m\sinh^3\!mL}.
$$
(Note that there is a factor of $L$ missing in the corresponding expressions
in {\it Class. Quant. Grav.} {\bf12} (1995) 1363.)
\vskip15truept
\noin{\bf Spin-half.}

I again show some corresponding spin-half results on cylinders and shells.

\noin For even $d$,
$$
\ze'_{I\times\rS^d}(0)=2^{(4-d)/2}\sum_{m=1}^\infty{1\over m}\big(
\cosech^d2mL-\cosech^dmL\big)
$$
while for odd $d$
$$\eqalign{
\ze'_{I\times\rS^d}(0)=&2^{(3-d)/2}\sum_{m=1}^\infty{1\over m}
\big(\cosech^d2mL-\cosech^dmL\big)\cr
&-2^{(d+3)/2}{L\over(d-1)!}\sum_{t=0}^{(d-1)/2} S^{(d)}_t\big
(2^{-2t-1}-1\big)\ze_R(-2t-1),\cr}
$$where the coefficients are defined by
$$\sum_{t=0}^{(d-1)/2} S^{(d)}_t x^t=\prod_{r=1}^{(d-1)/2}\big(x
-(r-1/2)^2\big).
$$

In particular
$$\eqalign{
&W_{\!I\times\rS^1}\!=\!-\!{1\over12}L
+\sum_{m=1}^\infty{1\over m}\big(\cosech 2mL-\cosech mL\big),\cr
\noalign{\vskip10truept}
&W_{\!I\times\rS^2}\!=\!\sum_{m=1}^\infty{1\over m}\big(\cosech^2 2mL-
\cosech^2 mL\big),\cr
\noalign{\vskip10truept}
&W_{I\times\rS^3}\!=\!{17\over480}L+{1\over2}\sum_{m=1}^\infty{1\over m}
\big(\cosech^3 2mL-\cosech^3 mL\big).\cr}
$$
Also
$$\eqalign{
&W_{2\shell}=\sum_{m=1}^\infty{2\over m}\bigg[\big(R^{2m}-R^{-2m}\big)^{-1}-
\big(R^m-R^{-m}\big)^{-1}\bigg]-{1\over6}\ln R
\cr
&W_{3\shell}=\sum_{m=1}^\infty{4\over m}\bigg[\big(R^{2m}-R^{-2m}\big)^{-2}-
\big(R^m-R^{-m}\big)^{-2}\bigg]+{1\over8}
\cr
&W_{4\shell}=\sum_{m=1}^\infty{4\over m}\bigg[\big(R^{2m}-R^{-2m}\big)^{-3}-
\big(R^m-R^{-m}\big)^{-3}\bigg]+{11\over180}\ln R\cr}
$$ where $R=r_1/r_2$ is the ratio of the radii of the shell.

We note that the coefficient of $\ln R$ is $\ze_{d-\ball}(0)$. This is valid,
up to the usual sign, for spin-zero as well and so, in the limit
$R\to\infty$, $W_{d-\shell}$ behaves like $W_{d-\ball}$ for a ball of
radius $R$ despite the
differing topologies.

If this result holds true for all $d$, it allows one to find $W_{d-\shell}$
for any even $d$ since $W_{\shell}-W_{\cylinder}$ in that case contains
only a $\ln R$ term whose coefficient can be found from the above arguments.
\vskip15truept
\noin{\bf Non-smooth boundaries.}

I now turn to another aspect of my talk. I wish to draw attention to
the fact that there exists a body of problems associated with manifolds whose
boundaries are only piecewise smooth. Such would be the case in simplicial
approximations.

In order to appreciate the geometry, it is easiest to picture a spherical
tetrahedron. This has an interior, $\man$, and a piecewise boundary,
$\pa\man$,
whose components are labelled $\pa\man_i$. These boundary components intersect
in {\it edges}, $E_{ij}$, which themselves intersect in {\it vertices},
$V_l$. We expect that the heat-kernel coefficients will contain contributions
from all these regions.

The $C_1$ coefficient is completely known so I will consider here $C_{3/2}$
which comes into play in three and higher dimensions and, hitherto, has not
beendiscussed.

The method of dimensions is sufficiently restrictive in this case to allow the
following general conjecture for the form of this coefficient,
$$\eqalign{
C^D_{3/2}=&{\sqrt\pi\over192}\!\sum_i\!\int_{\pa{\cal M}_i}\!\!
\big(6\tr(\ka_i^2)-\!3\ka_i^2-\!4\wR+12(8\xi-1)R\big)\cr
&-{\sqrt\pi\over24}\sum_{(ij)}\int_{E_{ij}}\bigg[
\la(\th_{ij})\,(\ka_i+\ka_j)+\mu(\th_{ij})\,(\ka^{(i)}+\ka^{(j)})\bigg]
+\sum_l\int_{V_l}\nu.\cr}
$$
$\ka_i$ is the extrinsic curvature of the codimension-1 boundary part
$\pa\man_i$, while $\ka^{(i)}$ is the extrinsic curvature of the edge
$E_{ij}$ considered as a codimension-1 submanifold of $\pa\man_i$.
$\th_{ij}$ is the dihedral angle along the edge $E_{ij}$. $\la$, $\mu$ and
$\nu$ are unknown functions.

Conformal invariance in three dimensions gives the relation
$$
2\tan(\th/2)\la(\th)+\mu(\th)=1
$$
and special case evaluation (the half-disc and cylinder) yields the
particular values,
$$
\la(\pi/2)=-3,\quad\mu(\pi/2)=7,
$$
which are in agreement with the general relation.
\vskip15truept
\noin{\bf Corner contributions.}

In three dimensions, $C_{3/2}$ is, up to a factor and zero modes, the constant
term in the heat-kernel expansion. I denote the
constant term coming from the vertex by
$w\big(\pi/\th_1,\pi/\th_2,\pi/\th_3\big)$ in the case of a trihedral vertex
with dihedral angles $\th_1,\th_2$ and $\th_3$.

The heat-kernel expansion on the right polygonal cylinder,
$I\times{\rm polygon}$, is easy to find and gives
$$
w\big({\pi\over\th},2,2\big)=\mp{1\over96}\bigg({\pi\over\th}-
{\th\over\pi}\bigg).
$$

Using the \zf\ on the fundamental domain $\fd$, it is straightforward to show
for the spherical tetrahedron that
$$\eqalign{
w(3,3,2)&=\mp1/16,\cr
w(3,4,2)&=\mp15/128,\cr
w(3,5,2)&=\mp15/64.\cr}
$$which gives us some information about the vertex function $\nu$.
\vskip15truept
\noin{\bf The smeared coefficient.}

For many purposes, it is convenient to take the distributional character
of the coefficient densities into account by introducing a smearing,
$$
C^{(d)}_k[g;f]\equiv\int_\man C^{(d)}_k(g,x,x)f(x).
$$
This can be determined from the conformal functional relation
$$
C^{(d)}_k\big[g;\de\om\big]=-{1\over d-k/2}\de C^{(d)}_k
\big[e^{-2\om}g;1\big]\big|_{\om=0}-2C^{(d)}_{k-1}[g;{\bf J}\de\om],
$$
with the operator
$$
{\bf J}=(d-1)\big(\xi-\xi(d)\big)\De_2,
$$
which follows, for example, by variation of the \zf. We find

$$\eqalign{
&C^{(d)}_{3/2}[g;f]=
{\sqrt\pi\over192}\sum_i\int_{\pa{\cal M}_i}\bigg[
\big(6\tr(\ka_i^2)-3\ka_i^2-4\wR\cr
&\hspace{***************}+12(8\xi-1)R\big)f
+30\ka n^\mu_if_\mu-24n^\mu_i n^\nu_i f_{\mu\nu}\bigg]\cr
&\hspace{**********}-{\sqrt\pi\over24}\sum_{(ij)}\int_{E_{ij}}\!\bigg[
\la(\th_{ij})\,(\ka_i+\ka_j)f
+\mu(\th_{ij})\,(\ka^{(i)}+\ka^{(j)})f\cr
&\hspace{***************}-{1\over2}\big(\mu(\th_{ij})+5\big)\,
(n^\mu_{(i)}+n^\mu_{(j)})f_\mu\bigg]+\sum_l\int_{V_l}\nu\,f,\cr}
$$ where
$f_\mu=\pa_\mu f$, $f_{\mu\nu}=f\cd{\mu\nu}$ and the $n^\mu$ are the various
normals.
\vskip15truept
\noin{\bf The cocycle function, $W[\bar g,g]$.}

The smeared coefficient is useful in calculating the connecting function
$W[\bar g,g]$.

In three dimensions, the conformal anomaly equation is
$$
\de W[\bar g]={1\over(4\pi)^{3/2}}\,C^{(3)}_{3/2}\big[\bar g;\de\om\big].
$$
After integration, this leads to
$$\eqalign{
W^D[\bar g,g]=&W^D_S[\bar g,g]
-\!{1\over384\pi}\!\sum_{(ij)}\!\int_{E_{ij}}\!\!
\bigg[2\big(\la(\ka_i\!+\!\ka_j)\!+\!\mu(\ka^{(i)}\!+\!\ka^{(j)})\big)\om
\bigg.\cr
&\bigg.-\!(5\!+\!\mu)\,(n^\mu_{(i)}\!+\!n^\mu_{(j)})\,\om_\mu
\!+\!4\om\big(n^\mu_{(i)}\!+\!n^\mu_{(j)}\big)
\om_\mu\bigg]+\sum_l\int_{V_l}\nu\,\om,\cr}
$$
where $W^D_S$ is the previous, smooth expression.
The effective action on some other manifolds can now be found, in particular
on those whose dihedral angles are $\pi/2$. (Otherwise we wouldn't know the
$\la$ and $\mu$ values.)

In this way the effective action on a 3-hemiball can be found
from that on a quarter 3-sphere and also that on a 3-hemishell from that on
the cylinder, $I\times$ 2-hemisphere. We find
$$\eqalign{
&W^D_{3\rm hemiball}-
W^D_{{1\over4}-3\rm sphere}={1\over384}(53-4\ln2)+{1\over48}\ln a,\cr
&W^D_{3\rm hemicap}-
W^D_{{1\over4}-3\rm sphere}=
{1\over96}\big(\ln(1-\cos\th)+8\cos\th+{21\over4}\cos^2\th\big),\cr
&W^D_{3\rm hemishell}-
W^D_{I\times2\rm hemisphere}=
{1\over96}\ln\big({r_1^3\over r_2}\big)+{9\over192},\cr
&W^D_{3\rm hemislice}-W^D_{3\rm hemishell}=
{1\over96}\bigg[2\ln(\cos\Th+\cos\De)-16\sin\Th\sin\De\cr
&\hspace{********************} +{21\over4}(\cos2\Th\cos2\De-1)\bigg].\cr}
$$
\vskip15truept
\noin{\bf The quartersphere effective action.}

In order to fix the determinant on the hemiball and hemicap, that on the
quarter 3-sphere is required. I again indicate how this is tackled. In this
case there are two perpendicular reflecting hyperplanes and all
the degrees are unity, except for $d_1=2$.

The \zf\ now is given by
$$
\ze_{QS}(s,a)=
\sum_{m,n=0}^\infty\comb{m+d-2}{d-2}{1\over{\big((a+2n+m)^2-\al^2\big)^s}}\,,
$$ where $\al=1/2$ for conformal coupling in
$d$-dimensions. Also $a=(d+3)/2$ for Dirichlet and
$a=(d-1)/2$ for Neumann conditions. This is a more difficult \zf\ to deal with
but the standard method of expanding in $\al$ yields after some algebra,
$$
\ze_{QS}'(0)=\ze_d'(0,d/2+1)+\ze_d'(0,d/2+2)
-\sum_{r=1}^u{1\over2^{2r}r}N_{2r}(d)\sum_{k=0}^{r-1}{1\over 2k+1},
$$where
$u$ is $d/2$ if $d$ is even, and $(d-1)/2$ if $d$ is odd.
$N$ is the residue of the Barnes \zf,
$$
\ze_d(s+r,a)\rightarrow{N_r(d)\over s}\quad{\rm as}\,\,s\rightarrow0,
$$where now
$$\eqalign{
\ze_d(s,a)&=
{i\Ga(1-s)\over2\pi}\int_L{e^{z(d/2+1/2-a)}(-z)^{s-1}\over
2^d\sinh^{d-1}(z/2)\sinh z}\,dz.\cr
\noalign{\vskip15truept}
&=\sum_{m,n=0}^\infty\comb{m+d-2}{d-2}{1\over{(a+2n+m)}^s}\,.\cr}
$$
$N$ is a generalised Bernoulli polynomial.

Rearrangement of the summation, as before, allows one to find that

$$
W^D_{{1\over4}-3\rm sphere}=-{1\over2}\ze'_{QS}(0)
={1\over4}\ze'_R(-2)-{1\over24}\ze'_R(-1)-{4\over3}\ln2-{1\over16}.
$$

The calculation can be performed for any dimension $d$ and also for the
case when there are $q$ hyperplanes inclined at $\pi/q$.

\vskip15truept
\noin{\bf B. Direct Method.}

I now turn to a brief discussion of the evaluation of determinants when the
eigenvalues are known only implicitly.

The method is based on the asymptotic behaviour of the quantity

$$
\ze(s,m^2)=\sum_\la{1\over(\la+m^2)^s},
$$
for large $m^2$. (This general approach in spectral geometry goes back a
long way to Carleman and to Dikii.)

The massive determinant is
$$
D(-m^2)=\exp\big(-\ze'(0,m^2)\big).
$$
One constructs the Weierstrassian product (equivalent to subtracting the
first $[d/2]$ terms in the Taylor expansion of $\ln(1+m^2/\la)$),
$$
\De(-m^2)=\prod_\la(1+{m^2\over\la})\exp\sum_{k=1}^{[d/2]}{1\over k}
\big({-m^2\over\la}\big)^{k},
$$
in terms of which it can be shown from a result of Voros that
$$
\ze'(0,0)=\lim_{m\to\infty}\ln\De(-m^2),
$$
so that all we have to do is to pick out the $m$-independent part of this
asymptotic limit.
\vskip15truept
\noin{\bf Determinant on the Euclidean ball.}

As an example I discuss the ball, for which some results are known by the
conformal method as recounted earlier.

The eigenvalues $\la=\al^2$ are the roots of
$$
J_p(\al)=0,\quad{(\rm Dirichlet).}
$$
and the degeneracy for a given Bessel order for even $d$ is
$$N^{(d)}_p={2\over(d-2)!}p^2(p^2-1)\ldots\big(p^2-(d/2-2)^2\big),
$$
and for odd $d$ :
$$N^{(d)}_p={2\over(d-2)!}p(p^2-1/4)\ldots\big(p^2-(d/2-2)^2\big).
$$

The Mittag-Leffler theorem allows one to write (this goes back to Euler)
$$
z^{-p}J_p(z)={1\over2^pp!}\prod_\al\bigg(1-{z^2\over\al^2}\bigg)
$$
or, setting $z=im$,
$$
m^{-p}I_p(m)={1\over2^pp!}\prod_\al\bigg(1+{m^2\over\al^2}\bigg).
$$

Moss has used this method to deduce the heat-kernel coefficients. D'Eath and
Esposito have also used this approach in calculations of, for example, the
conformal anomaly on balls. We are here interested in the functional
determinant.
\vskip15truept
\noin{\bf The 4-ball.}

The particular case of the 4-ball illustrates the technique. We have
$$\eqalign{
&\De(-m^2)=
\prod_{p,\al_p}(1+{m^2\over\al_p^2})\exp
\big(-{m^2\over\al_p^2}+{m^4\over2\al_p^4}\big)\cr
&=\prod_p\big(p!2^pm^{-p}I_p\big)\exp\big(-
{m^2\over4(1+p)}+{m^4\over32(1+p)^2(2+p)}\big)\cr}
$$
where Rayleigh's formulae for the sums of inverse powers of roots of the Bessel
function have been employed. (Actually these are not needed as they yield
mass-{\it dependent} terms in the limit and can be ignored.)

The asymptotic expansion of $I_p$ has been derived by Olver and is easily
found using computer algebraic manipulation from recursion relations. Then,
$$\eqalign{
\ln\De(-m^2)\sim&
\sum_{p=1}^\infty p^2\bigg(p\ln2+\ln p!-\ln\sqrt2\pi+\ep-p\ln(p+\ep)\cr
&-\ln\sqrt\ep+\sum_n{T_n(t)\over \ep^n}
-{m^2\over4(1+p)}+{m^4\over32(1+p)^2(2+p)}\bigg),\cr}
$$
where $\ep^2=p^2+m^2$, $t=\ep/m$ and the polynomials $T_n(t)$ come from a
cumulant expansion of Olver's asymptotic series.

I would like to take you through all the technicalities but this is not the
place. I will just say that an integral representation for
$\ln p!$ is introduced and the asymptotic lmits of all the summations
found by repeated use of the Watson-Kober formula,
$$\eqalign{
\sum_{p=1}^\infty\bigg({1\over(p^2+m^2)^s}-\sum_{n=0}^{M-1}\comb{-s}n
{m^{2n}\over p^{2s+2n}}\bigg)&={\sqrt\pi\Ga\big(s-1/2\big)
m^{1-2s}\over2\Ga(s)}-{1\over2}m^{-2s}\cr
&+{2\pi^s\over\Ga(s)}m^{(1-2s)/2}\sum_{p=1}^\infty p^{(2s-1)/2}
K_{(2s-1)/2}(2\pi mp)\cr
&\hspace{*****}-\sum_{n=0}^{M-1}\comb{-s}n m^{2n}
\ze_R(2s+2n),\cr}
$$
for $\Real s>1/2-M$,
producing some amusing identities and relations along the way.

The result for the 4-ball agrees with that from the conformal method so I
content myself with exposing the 6-ball formula,
$$
\ze_6'(0)=-{4027\over6486480}-{1\over756}\ln2+{1\over60}\ze_R'(-5)
-{1\over24}\ze_R'(-4)+{1\over24}\ze_R'(-2)-{1\over60}\ze_R'(-1),
$$
derived, independently, by Bordag, Geyer, Elizalde and Kirsten recently, using
a comparable, but different in detail, method.

For the $d$-ball, all terms except the fractional part are easily deduced.
\vskip15truept
\noin{\bf Comments.}

This talk has simply been a litany of results with
no motivation and it is not my intention to change this now.
However a few historical comments are quite in keeping.

Substantial relevant calculations have been undertaken, mostly in the context
of quantum cosmology. In particular D'Eath and Esposito have calculated,
amongst other things, the conformal anomaly, $\ze(0)$, for massless spinor
fields on the 4-ball. Their value of $11/180$ can be seen in our expression
for $W_{\rm 4cap}$. Since, in this case, $\ze(0)$ is a conformal invariant,
it is much easier to evaluate it on the hemisphere. The same quantity has been
calculated by Kamenshchik and Mishakov using the technique developed by
Barvinsky, Kamenshchik, Karmazov and Mishakov, which is another means of
extracting the asymptotic behaviour of the eigenfunctions, although some
use of Olver's expansions still seems necessary. $\ze(0)$ is calculated on a
cap of the 4-sphere for massive spinors, and a limit taken to flat space.

As is well appreciated, $\ze(0)$ also follows directly from the expression
for the appropriate heat-kernel coefficient, when this is known and relevant.

Kamenshik and Mishakov also calculate that $\ze(0)=0$ for massless
fermions on the 4-shell. In fact this vanishing is true for all shells and is
geometrically obvious for even dimensions but is a result of calculation for
odd dimensions.

The theory of gauge fields is these situations is a separate chapter. See
the contributions by Esposito, Moss and Vassilevich to this Seminar.

The computations of the corresponding functional determinants in these
geometries are less numerous than those of the conformal anomalies. I will
attempt a survey in another place.

\bye